\documentclass[twoside]{article}

\usepackage[sc]{mathpazo}
\usepackage[T1]{fontenc}
\usepackage{microtype}

\usepackage{amssymb,amsmath}
\usepackage{graphicx}
\usepackage{color}

\usepackage{float}
\usepackage{psfrag}
\newcommand{\V}[2]{V^{#1#2}}
\newcommand{\abs}[1]{\left| #1 \right|}
\newcommand{\nPartTypes}{K}

\usepackage[hmarginratio=1:1,top=32mm,columnsep=20pt]{geometry}
\usepackage[font=it]{caption}
\usepackage{paralist}
\usepackage{multicol}

\usepackage{lettrine}

\usepackage{abstract}

\usepackage{titlesec}
\renewcommand\thesection{\Roman{section}}
\titleformat{\section}[block]{\large\scshape\centering}{\thesection.}{1em}{}

\usepackage{fancyhdr}
\usepackage{hyperref}

\title{\vspace{-15mm}%
	\fontsize{24pt}{10pt}\selectfont
	\textbf{A Design Path for Hierarchical Self-Assembly of Patchy Colloids}
	}	
\author{%
	\large
	\textsc{E. Edlund, O. Lindgren, M. Nilsson Jacobi}  \\[2mm]%\thanks{Template by \href{http://www.howtotex.com}{howtoTeX.com}}	
	\normalsize	Chalmers University of Technology, Gothenburg \\
	\normalsize	mjacobi@chalmers.se
	\vspace{-5mm}
	}
\date{}

\begin{document}

\maketitle
\thispagestyle{fancy}

\begin{abstract}
\noindent Patchy colloids are promising candidates for building blocks in directed self-assembly. To be successful the surface patterns need to both be simple enough to be synthesized, while feature-rich enough to cause the colloids to self-assemble into desired structures. Achieving this is a challenge for traditional synthesis methods. Recently it has been suggested that the surface pattern themselves can be made to self-assemble. In this paper we show that a wide range of functional structures can be made to self-assemble using this approach. More generally we present a design path for hierarchical targeted self-assembly of patchy colloids. At the level of the surface structure, we use a predictive method utilizing universality of patterns of stripes and spots, coupled with stoichiometric constraints, to cause highly specific and functional patterns to self-assemble on spherical surfaces.  We use a minimalistic model of an alkanethiol on gold as a model system and demonstrate that, even with limited control over the interaction between surface constituents, we can obtain patterns that causes the colloids themselves to self-assemble into various complex geometric structures. We demonstrate how variations of the same design path cause in-silico self-assembly of strings, membranes, cubic and spherical aggregates, as well as various crystalline patterns.

\end{abstract}

%%%%%%%%%%%%%%%%%%%%%%%%%%%%%%%%%%%%%%%%%%

\begin{multicols}{2}
\lettrine[nindent=0em,lines=3]{D}uring the last decade, patchy colloids have emerged as candidate building blocks for colloidal self-assembly. By altering the position of attractive patches, patchy colloids can for example be used to mimic the anisotropic interactions between folded proteins, drawing inspiration from how certain proteins self-assemble into capsid shells enclosing the viral genome \cite{Gira_review_2013}. With an ambition towards self-assembling meta materials and nano sized functional structures, the number of colloidal structures self-assembled in-silico or in-vivo have since then grown to encompass structures ranging from kagome lattice systems to colloidal strings \cite{Chen_Kagome_2011, Smallenburg_strings_2012} using equally varied techniques. In this paper, we demonstrate in-silico hierarchical self-assembly of a range of geometrical structures: different crystals, membranes, strings, vesicles of different size and cubic aggregates, all in the context of a single model system. By modifying parameters of the colloids' coating, we are able to select which of the geometries that self-assembles. The parameter choices are based on a combination of more general design principles, also presented in this paper, and analytical predictions of the self-assembly processes \cite{Edlund_Predicting_2014}.

%Inspired in part from how certain proteins self-assembles into capsid shells to protect the viral genome that encodes them, patchy particles have been designed to show similar self-assembling behavior. In the latter case by using the orientation of the patches as a representation of the anisotropic interactions between folded proteins [CITE].

Synthesization of large quantities of colloids with attractive patches in specific directions and high specificity is difficult. The methods developed for this in the last decade \cite{Pawar_fabrication_2010} range from various coating techniques on thin films of colloids on a substrate like dip-coating and glancing angle deposition \cite{Pawar_Glancing_2009}, to techniques where colloids are aligned using an external electric field combined with an evaporator \cite{Arnaud_Networks_2014}, to multiphase colloidal particles where constituents are added or subtracted sequentially to produce highly anisotropic colloidal shapes \cite{sacanna13}. A rather different method, suggested recently, is to let the surface coating itself self-assemble into desired patterns as a part of a hierarchical self-assembly process. Experimental findings indicate that certain mixtures of alkanethiols adsorbed on gold nanoparticles phase separate into various morphologies depending on the properties of ligands and nanoparticles~\cite{jackson_spontaneous_2004}. The results in this article support the idea of self-assembled coatings on patchy colloids by showing that limited control over how the ligands interact is sufficient to template for targeted self-assembly of a wide range of surface patterns, and in extension, colloidal structures. Letting the surface patterns self-organize naturally leads to patches and stripes which are well suited for directing self-assembly on the colloidal scale. An important advantage of using self-assembly to form the patterns is that there is no intrinsic difficulty in making patterns with many patches and high specificity. The method can therefore give access to many patterns that seem hard to come by using other techniques.

%%%%%%%%%%%%%%%%%%%%%%%%%%%%%%%%%%%%%%%%%%%%%%%%%%%%%%%%%%%%%%%%

\section{A model system --- alkanethiol-on-gold}

A proposed candidate for self-assembling patchy morphologies on spheres is the alkanethiol-on-gold system~\cite{jackson_spontaneous_2004}. 
It consists of spherical gold nanoparticles (colloids) coated with different species of alkanethiols. 
These alkanethiols organize into patterns due to a competition between their immiscibility and a mixing effect of entropic origin~\cite{singh_entropy-mediated_2007}. 
Depending on the number of alkanethiol types, many different patterns can appear~\cite{pons-siepermann_design_2012a,pons-siepermann_design_2012b}.

Here we use a simplified model of the alkanethiol-on-gold system, described in detail in \cite{Edlund_Predicting_2014}. 
It models the alkanethiols as spherical particles interacting with a set of effective interactions. 
Between alkanethiols of different type, the interaction consists of a hard core with diameter $\sigma_0$, a soft shoulder potential with range $\sigma_1$ (causing immiscibility), and a square-well potential of depth $\epsilon$ (representing the mixing effect),
\begin{equation}
\tag{\addtocounter{equation}{1}\arabic{equation}a}
\label{SAM-model_a}
\V{\alpha}{\beta}(r)= \begin{cases} 
	\infty, & \mbox{if } r < \sigma_0 \\ 
	1, & \mbox{if } \sigma_0 < r <  \sigma_1 \\ 
	-\epsilon, & \mbox{if } \sigma_1 < r <  \abs{L_\alpha - L_\beta} \\ 
	 0, & \mbox{otherwise}
	\end{cases},
\end{equation}
where the $L_\alpha$ are abstracts parameters defined for each alkanethiol type, representing the different lengths or bulkiness of the tail groups. 
The interaction between alkanethiols of the same type is taken to be a simple hard-core repulsion.
\begin{equation}
\tag{1b}\label{SAM-model_b}
\V{\alpha}{\alpha}(r)= \begin{cases} 
	\infty, & \mbox{if } r < \sigma_0 \\
	 0, & \mbox{otherwise}
	\end{cases}.
\end{equation}

While minimalistic, it captures much of the behavior of  the original system~\cite{Edlund_Predicting_2014}. 
It is also a good model system for our design-path since  the interactions are determined by only a few parameters, making the model more relevant as a proxy for an experimental system than a model with more complex interactions.
%We will thus use it as stand-in for an experimental system, its parameters to be determined for self-assembly of various morphologies.

%%%%%%%%%%%%%%%%%%%%%%%%%%%%%%%%%%%%%%%%%%%%%%%%%%%%%%%%%%%%%%%%

\section{Theory}

In~\cite{Edlund_Predicting_2014} % we introduced a method for predicting the morphology of self-assembling systems given an effective model of its interactions. We will here give a short summary of the theory.
we formulate the problem of predicting the morphology of a particle system in terms of a Potts-like spin model \cite{baxter_exactly_1982}. Assuming that we have a model of the effective interactions in the systems, here described as a set of potentials $V^{\alpha\beta}(r)$, we can construct a Hamiltonian for this spin model as 
\begin{eqnarray} \label{hamiltonian}
H = \sum_{\alpha \beta}^\nPartTypes \sum_{ij}^N \Pi_{i\alpha} \, \V{\alpha}{\beta}_{ij} \, \Pi_{j\beta}
 \end{eqnarray}
 where $N$ is the number of lattice sites, $K$ the number of particle types, $\V{\alpha}{\beta}_{ij} = \V{\alpha}{\beta}(\left| \vec{r_i} - \vec{r_j} \right|)$, and $\Pi_{i\alpha}$ is 1 if site $i$ contains a particle of type $\alpha$ and 0 otherwise. 

The task is then to find the low-energy states of this Hamiltonian. While we cannot hope to solve it exactly for general interaction matrices $V$, if we relax the constraints and allow the state $\Pi$ to take any real values while keeping its norm fixed we get a quadratic problem that is solvable. First the potentials $V^{\alpha\beta}$ are diagonalized independently by using the spherical harmonics analogue of the Fourier transform, then one small matrix for each orbital number is diagonalized. The result is an energy spectrum with $K$ branches whose minimum will be the ground state for the relaxed model.

\begin{figure*}
%	\psfrag{a}[c][c][\psLabelSize][0]{(b)}
%	\psfrag{b}[c][c][\psLabelSize][0]{(a)}
%	\psfrag{c}[c][c][\psLabelSize][0]{(c)}
%	\psfrag{E}[c][c][\psAxesLabelSize][0]{$E$}
%	\psfrag{l}[c][c][\psAxesLabelSize][0]{$l$}
%	\psfrag{1}[c][c][\psTickSize][0]{$1$}
%	\psfrag{4}[c][c][\psTickSize][0]{$4$}
%	\psfrag{5}[c][c][\psTickSize][0]{$5$}
%	\psfrag{t}[cr][c][\psTickSize][0]{$-10$}
%	\psfrag{T}[cr][c][\psTickSize][0]{$10$}
%\includegraphics[width= 0.31\textwidth]{simpleSpectrumC}
%\includegraphics[width= 0.31\textwidth]{exampleB}
%\includegraphics[width= 0.31\textwidth]{exampleC}
\includegraphics[width= 0.93\textwidth]{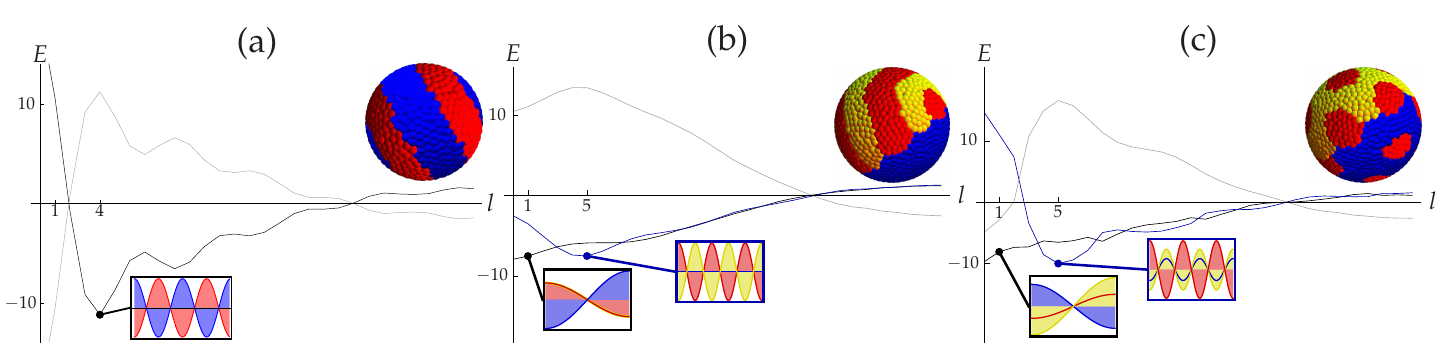}
\caption{ \label{simpleMin} The length scale that defines the segregation between different surface constituent types is given by the energy spectra. To fully understand the ground state morphology, stoichiometric constraints has to be taken into account as well. a) With two particle types of equal concentration, the ground state will be stripes. The wavelength is determined by the minimum in the energy spectrum. Here $L_{blue}-L_{red} = 1.1$ radii (radius of the colloid), $\epsilon = 0.1$. b) With multiple particle types, the ground state is described by multiple minima in the energy spectrum, sufficiently many to account for the entire stoichiometry. The global minimum takes precedence and separates blue from red and yellow, the second minimum then determines the wavelength of the yellow-red striped part.  $L_{blue} = 0.44$, $L_{red} = 0$, $L_{yellow} = 0.88$ radii, $\epsilon = 0.1$.  
c) Here the dominating segregation is between red and the surrounding colors, but stripes cannot form due to stoichiometric constraints (red, blue and yellow have equal proportions), causing the red state to be spotty instead. Here $L_{blue} = 1.1$, $L_{red} = 0$, $L_{yellow} = 1$ radii, $\epsilon = 0.1$.
}
\end{figure*}

Each branch describes variation between two partitions of the particle types. To predict the behavior of the original model we combine the minima of enough branches to completely specify a particle configuration, excluding unphysical branches that does not describe separations between any types. An example is shown in Fig.~\ref{simpleMin} B, where the global minimum describes the phase separation of the blue particles and the next branch is needed to determine the behavior of the red and yellow particles, here a striped state. For further details see \cite{Edlund_Predicting_2014}.

\section{A design path for targeted hierarchical self-assembly of patchy nano-particles}

Since the self-assembling morphologies are so closely tied to their energy spectra, we can use it as an intermediate between the parameters describing the coating and the resulting ground state. The theoretical model for what constitutes a ground state helps us decide on reasonable targets to aim for, patterns with stripes and spots symmetrically distributed over the surface, and which corresponding features are required in the energy spectrum. %The theory also significantly speed up the design process as testing a set of interactions by simulations or experimentally takes a much longer time than performing the spectral analysis on the same set of parameter. %In addition this analysis helps with the identification of interaction parameters that are important for the pattern formation.

While different functional patterns require different approaches, the overarching design path can be summarized in a few steps;

\begin{compactitem}
\item The locations of attractive patches in the target surface pattern are determined by the target nano-particle configuration. 
\item An achievable target pattern is one with a hierarchy of spots and stripes symmetrically distributed over the surface.
\item Decompose the target pattern into spherical harmonics. The weights on the harmonic modes identify the length scales at which segregation occurs and which segregations takes precedence.
\item Iterate through different parameter settings for the surface constituents, until the effective interaction potentials transforms into an energy spectrum with the necessary features.
\item Set the stoichiometry to match the target pattern.
\item Verify in-silico that the designed interactions lead to the desired surface pattern.
\end{compactitem}

If the stoichiometry is too skewed in the attempt to create patches with high specificity, the ground state can turn out to be a pattern of fewer, clustered patches instead. The following sections cover a straight forward example of the design path followed by examples of what is possible given different degree of control over the energy spectrum and the interactions between surface constituents.

%%%%%%%%%%%%%%%%%%%%%%%%%%%%%%%%%%%%%%%%%%%%%%%%%%%%%%%%%%%%%%%%
\section{Single length scale coupled with stoichiometric constraints}

While the square well potentials of the model (Eqs.~\ref{SAM-model_a} and b) are not realistic, striped patterns have been observed both in first principle simulations \cite{pons-siepermann_design_2012a} and experiments \cite{Godin_alkanethiol_2004}.
There are also fundamental theoretical reasons explaining why stripes are expected to appear universally \cite{Edlund_stripes_2010}. This supports the idea that at least one (non $l=0$) minimum in the energy spectrum can be selected for, the minimum that determines the length scale of the striped pattern. This is sufficient for causing hierarchical self-assembly of different crystal patterns when combined with stoichiometric constraints.

\begin{figure*}[htb]
%	\psfrag{a}[c][c][\psLabelSize][0]{(a)}
%	\psfrag{b}[c][c][\psLabelSize][0]{(b)}
%	\psfrag{c}[c][c][\psLabelSize][0]{(c)}
%	\psfrag{E}[c][c][\psAxesLabelSize][0]{$E$}
%	\psfrag{l}[c][c][\psAxesLabelSize][0]{$l$}
%	\psfrag{1}[c][c][\psTickSize][0]{$1$}
%	\psfrag{3}[c][c][\psTickSize][0]{$3$}
%	\psfrag{4}[c][c][\psTickSize][0]{$4$}
%	\psfrag{t}[cr][c][\psTickSize][0]{$-10$}
%	\psfrag{T}[cr][c][\psTickSize][0]{$10$}
%\includegraphics[width= 0.31\textwidth]{latticeSpectrum}
%\includegraphics[width= 0.35\textwidth]{diamondLattice}
%\includegraphics[width= 0.27\textwidth]{cubicLattice}
\includegraphics[width= 0.93\textwidth]{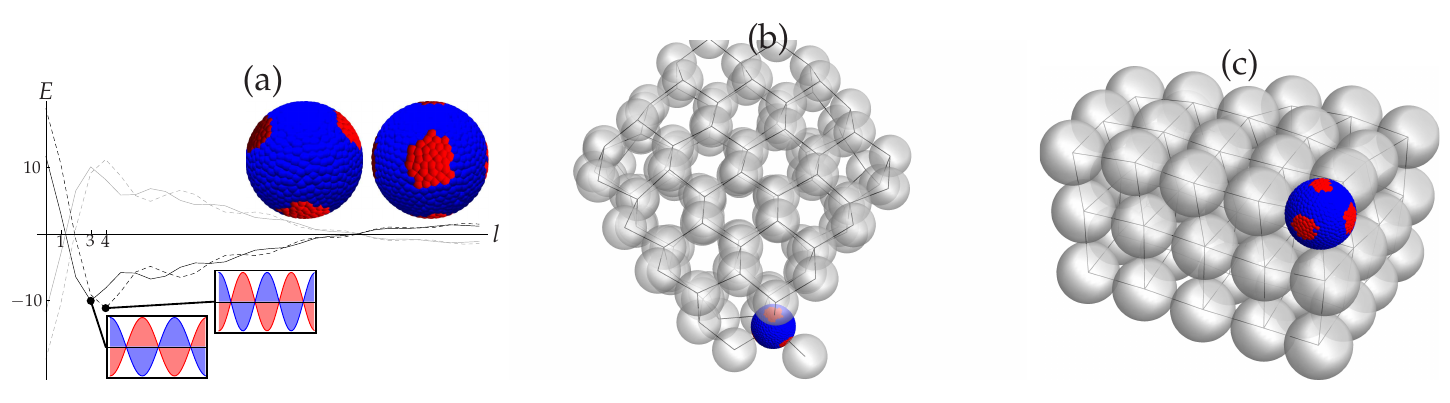}
\caption{ \label{latticeAssembly} 
a) Phase separation at $l=3$ together with a 4:1 ratio of blue:red surface constituents causes the ground state configuration to have four equally separated patches (solid line, left inset in a), $L_{red}-L_{blue} = 1.33$ radii, $\epsilon = 0.05$). With the minimum at $l = 4$ and a blue:red ratio of 3:1, the coating self-assembles into 6 patches instead (dashed line, right inset in a), $L_{red}-L_{blue} = 1$ radius, $\epsilon = 0.1$). On the colloidal scale, the small changes makes the difference between a self-assembling diamond lattice b) or a cubic lattice c). Both surface patterns and the colloidal structures are obtained from Monte Carlo simulations using simulated annealing. 
}
\end{figure*}

Suppose that we want to cause patchy particles to self-assemble into a crystal pattern, for example a diamond or a cubic lattice. 
A simple pattern that achieve this would be patches in each direction where a neighboring colloid should reside. For a diamond lattice, this would mean four patches equally distanced to each other on the surface of the nanoparticle, for a cubic lattice, the number of patches would be six.  Expanding these patterns in terms of spherical harmonic functions shows that they are almost completely dominated by $l = 0$ modes, determining the stoichiometry, as well as $l = 3$ resp. $l = 4$ modes, determining the length scale of the spotted pattern. By choosing parameters in the alkanethiol-on-gold model (or indeed, in an experimental setup) that causes the global minimum of the energy spectrum, as well as the stoichiometry, to coincide with the target pattern's will cause the target pattern to self-assemble, see Fig.~\ref{latticeAssembly} A. Systems of particles with attractive patches positioned according to the self-assembling surface patterns will at the next level self-assemble into desired colloidal lattices according to our simulations, Fig.~\ref{latticeAssembly} B-C. 

%[Does this really fit here?] Target patterns where patches are equally spaced over the surface are easier to realise, since the periodicity of the surface partially slaves the distances between patches. On the other hand, if the stoichiometry becomes too skewed in the attempt to create patches with higher specificity, the ground state can turn out to be a pattern of fewer, larger patches clustered to one side of the colloid. See Fig.~\ref{cubeFig}  or fig.~\ref{stringMembraneAssembly} for an example of how to circumvent this limitation on patch specificity.

%If the patches in the target pattern have too high specificity, the ground state could end up being one with fewer patches instead. This starts to be a problem when the target pattern becomes less localised to one orbital number with the introduction of overtones etc. Similarly, a target pattern where for example patches are equally spaced over the surface is more easily realised, since the periodicity of the surface partially slaves the distances between patches.
 
%%%%%%%%%%%%%%%%%%%%%%%%%%%%%%%%%%%%%%%%%%%%%%%%%%%%%%%%%%%%%%%%

\section{Single length scale coupled with stoichiometric constraints and segregating subtypes}

Using only two particle types, the patterns tend to be spherical symmetric and the previous approach of selecting a single length scale usually falls short when trying to obtain more complicated (asymmetric) patterns. This issue can be remedied by introducing more surface constituent types. For simplicity, lets consider the introduction of very similar subtypes of one particle type; type $blue$ is replaced by type $blue$ and $yellow$. These subtypes behaves identical at long range ($L_{blue} = L_{yellow}$), but they are immiscible and tend to segregate, all else being equal.
 The internal segregation between subtypes causes the system to strive towards achieve something similar to the previous striped or spotted pattern, while also minimizing the subtype $blue$ to $yellow$ interfaces. If we choose a stoichiometry where the subtypes can completely occupy different subsets of the previously $blue$ spots/stripes, we can completely retain the ground state pattern, without any subtype interface. Such a pattern should be a ground state. 

\begin{figure*}[htb]
%	\psfrag{a}[c][c][\psLabelSize][0]{(a)}
%	\psfrag{b}[c][c][\psLabelSize][0]{(b)}
%	\psfrag{c}[c][c][\psLabelSize][0]{(c)}
%	\psfrag{E}[c][c][\psAxesLabelSize][0]{$E$}
%	\psfrag{l}[c][c][\psAxesLabelSize][0]{$l$}
%	\psfrag{1}[c][c][\psTickSize][0]{$1$}
%	\psfrag{4}[c][c][\psTickSize][0]{$4$}
%	\psfrag{t}[cr][c][\psTickSize][0]{$-10$}
%	\psfrag{T}[cr][c][\psTickSize][0]{$10$}
%\includegraphics[width= 0.31\textwidth]{membraneStringSpectrum}
%\includegraphics[width= 0.31\textwidth]{stringConfig}
%\includegraphics[width= 0.31\textwidth]{membraneConfig}
\includegraphics[width= 0.93\textwidth]{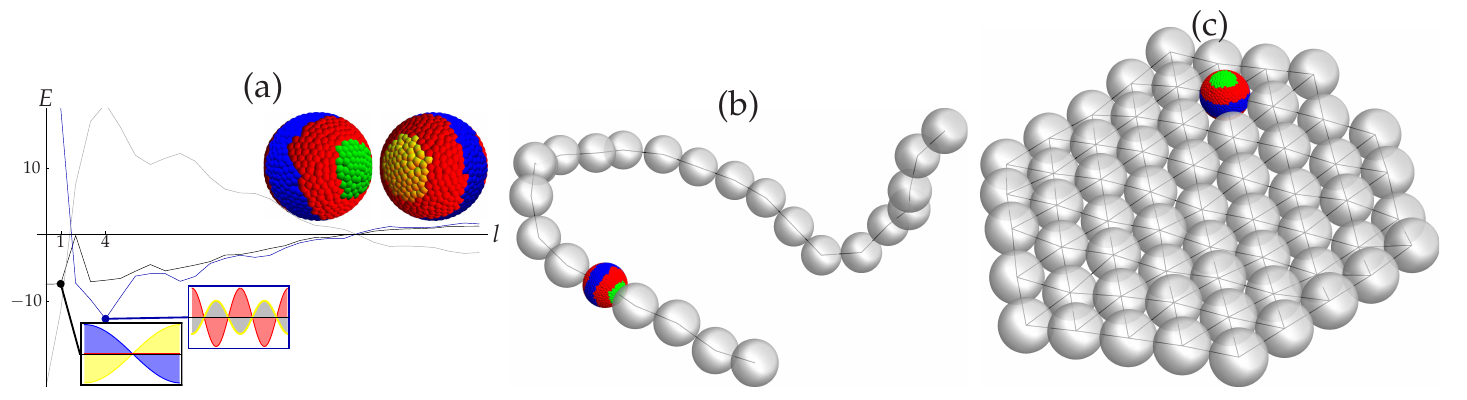}
\caption{ \label{stringMembraneAssembly} 
a) The global minimum causes segregation into two red stripes. The other three colors, being immiscible, fills in different segments caused by the red stripes. The inset in a) shows the self-assembled patchy particle from different directions. Depending on which segments are made attractive, the colloidal system self-assembles into different structures. With opposing ends attractive, we obtain strings b) and with an attractive central stripe, we obtain a self-assembled membrane c) when testing the interactions Monte Carlo simulations ($L_{blue} = L_{yellow} = L_{green} = L_{red}-1.1$ radii, $\epsilon = 0.1$).}
\end{figure*}

We now show how this can be used to design hierarchical self-assembly of strings and membranes. Patterns with two patches at opposing sides of every nanoparticle would cause the self-assembly of a string, while a single attractive band at the equatorial would cause the self-assembly of a membrane. One way to select for this pattern would be to choose parameters for a binary coating where segregation occurs at length scale corresponding to $l = 4$, causing a striped pattern as the ground state. By differentiating type $blue$ into subtypes $blue$, $yellow$, and $green$,  we can ensure that e.g. $yellow$ and $green$ occupy each polar patch respectively, and $blue$ occupies the equatorial stripe. If $yellow$ and $green$ are attractive, the polar patches will be too, causing self-assembly of strings. If $blue$ is attractive, the nanoparticles will instead form membranes. Fig.~\ref{stringMembraneAssembly} shows how this construction cause hierarchical self-assembly of string and membranes. %particles with patterns of positive patches (and stripes) obtained from Monte Carlo annealing of the surface configurations themselves.

%%%%%%%%%%%%%%%%%%%%%%%%%%%%%%%%%%%%%%%%%%%%%%%%%%%%%%%%%%%%%%%%

\section{Various degree of immiscibility coupled with stoichiometric constraints}

By controlling the strength of the energetic drive towards segregation, we can also cause hierarchical self-assembly of vesicles. A thin, circular, attractive band causes colloids to self-assemble into vesicles, where the distance from the equatorial to the stripe determines the curvature of the vesicles. If the energy associated to the interface between a type $red$ and a type $blue$ is twice as large as the interfaces with type $yellow$, the $yellow$ constituent will form a stripe separating the others. The features are represented in the energy spectrum as a large minimum at a low $l$ value, forcing two types to segregate to opposing sides and a much smaller minimum at the length scale corresponding to the width of the band that separates the two sides, see Fig.~\ref{stringMembraneAssembly}. Since the low $l$ minimum is an order of magnitude larger than the second minimum, the ground state configuration will be one where type $blue$ and $red$ are separated as much as possible. This differentiates the target pattern from e.g. one where $blue$ and $red$ forms a Janus sphere with $yellow$ as an overlapping spotty pattern as can be seen in Fig.~\ref{latticeAssembly} C. The position of the attractive $yellow$ stripe is only dependent on the stoichiometry and by adjusting the ratio between the surface constituents we can choose between vesicles with different curvatures, see Fig.~\ref{stringMembraneAssembly}.

\begin{figure*}[htb]
%	\psfrag{a}[c][c][\psLabelSize][0]{(a)}
%	\psfrag{b}[c][c][\psLabelSize][0]{(b)}
%	\psfrag{c}[c][c][\psLabelSize][0]{(c)}
%	\psfrag{E}[c][c][\psAxesLabelSize][0]{$E$}
%	\psfrag{l}[c][c][\psAxesLabelSize][0]{$l$}
%	\psfrag{1}[c][c][\psTickSize][0]{$1$}
%	\psfrag{7}[c][c][\psTickSize][0]{$7$}
%	\psfrag{t}[cr][c][\psTickSize][0]{$-10$}
%	\psfrag{T}[cr][c][\psTickSize][0]{$10$}
%\includegraphics[width= 0.31\textwidth]{vesiclesSpectrum}
%\includegraphics[width= 0.31\textwidth]{icoConfig}
%\includegraphics[width= 0.31\textwidth]{largeVescConfig}
\includegraphics[width= 0.93\textwidth]{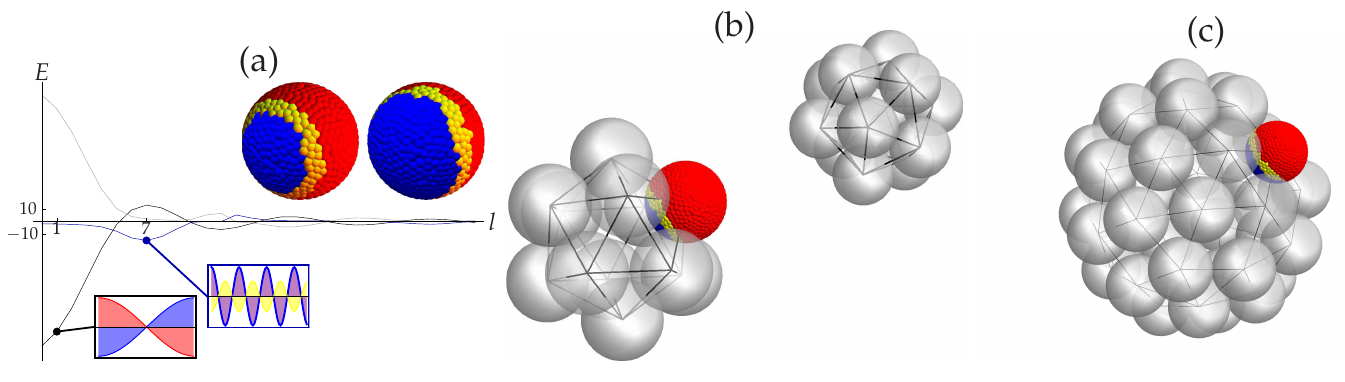}
\caption{ \label{lengthfig} 
a) The dominating global minimum separates red and blue to opposing sides of the colloid. The second minimum  at $l = 7$ encourages an even yellow stripe of a thickness determined by the stoichiometry. By adjusting the stoichiometry, the distance to the yellow stripe and the equatorial can be tuned. b) A large distance from stripe to the equatorial will cause an attractive yellow stripe to force self-assembly of small vesicles. c) A smaller distance will cause the vesicle surface curvature to be lower and the vesicle(s) to be larger. All cross color particle interactions are shoulder potentials, but interactions with $yellow$ particles are weaker (by a factor of $0.4$).}
\end{figure*}

%%%%%%%%%%%%%%%%%%%%%%%%%%%%%%%%%%%%%%%%%%%%%%%%%%%%%%%%%%%%%%%%

\section{Multiple minima in the energy spectrum}

We identify parameters in the alkanethiol-on-gold model that cause self-assembly into colloidal particles with three attractive patches separated by $90^\circ$, which will cause the colloids to self-assemble into cubes. It turns out that this target requires a more complicated construction than the previous. The patches of the desired pattern are primarily described by $l=4$ modes. With two polymer types, $L_{red}-L_{blue} = 1.$ radii, the surface pattern will form 6 patches separated by $90^\circ$ for stoichiometry 7:1 to 2:1. To only obtain 3 patches, a third type is introduced with the intention of creating a Janus sphere with the patches on one of the two sides. The third polymer length is chosen so that the energy spectrum exhibits a minimum at $l = 0$ %or $1$ 
that separates the third type from the previous two. An important property of the pattern is the specificity of the attractive patches. If the three red patches are small to increase the specificity on the interactions, this prevents other geometric structures with bonds in similar directions to assemble. This suggests a third minimum that also separates $blue$ and $red$ from $yellow$ at approximately double frequency $l = 2 \times 4$. The yellow regions that separates $blue$ and $red$ makes those patches thinner/smaller. If we try to reduce the patch sizes by stoichiometry alone, i.e.\ small red patches separated by a large yellow region, other patterns described by fewer components of the basis tend to form instead.

\begin{figure}[H] %Does not float, we might need to position this at a suitable position in the text later
%	\psfrag{a}[c][c][\psLabelSize][0]{(a)}	
%	\psfrag{b}[c][c][\psLabelSize][0]{(b)}
%	\psfrag{c}[c][c][\psLabelSize][0]{(c)}
%	\psfrag{E}[c][c][\psAxesLabelSize][0]{$E$}
%	\psfrag{l}[c][c][\psAxesLabelSize][0]{$l$}
%	\psfrag{1}[c][c][\psTickSize][0]{$1$}
%	\psfrag{4}[c][c][\psTickSize][0]{$4$}
%	\psfrag{7}[c][c][\psTickSize][0]{$7$}
%	\psfrag{t}[cr][c][\psTickSize][0]{$-10$}
%	\psfrag{T}[cr][c][\psTickSize][0]{$10$}
%\centering\includegraphics[width= 0.40\textwidth]{AggCubeSpectrum}
%\newline
%\centering\includegraphics[width= 0.40\textwidth]{cubicAggregate}
\centering\includegraphics[width= 0.40\textwidth]{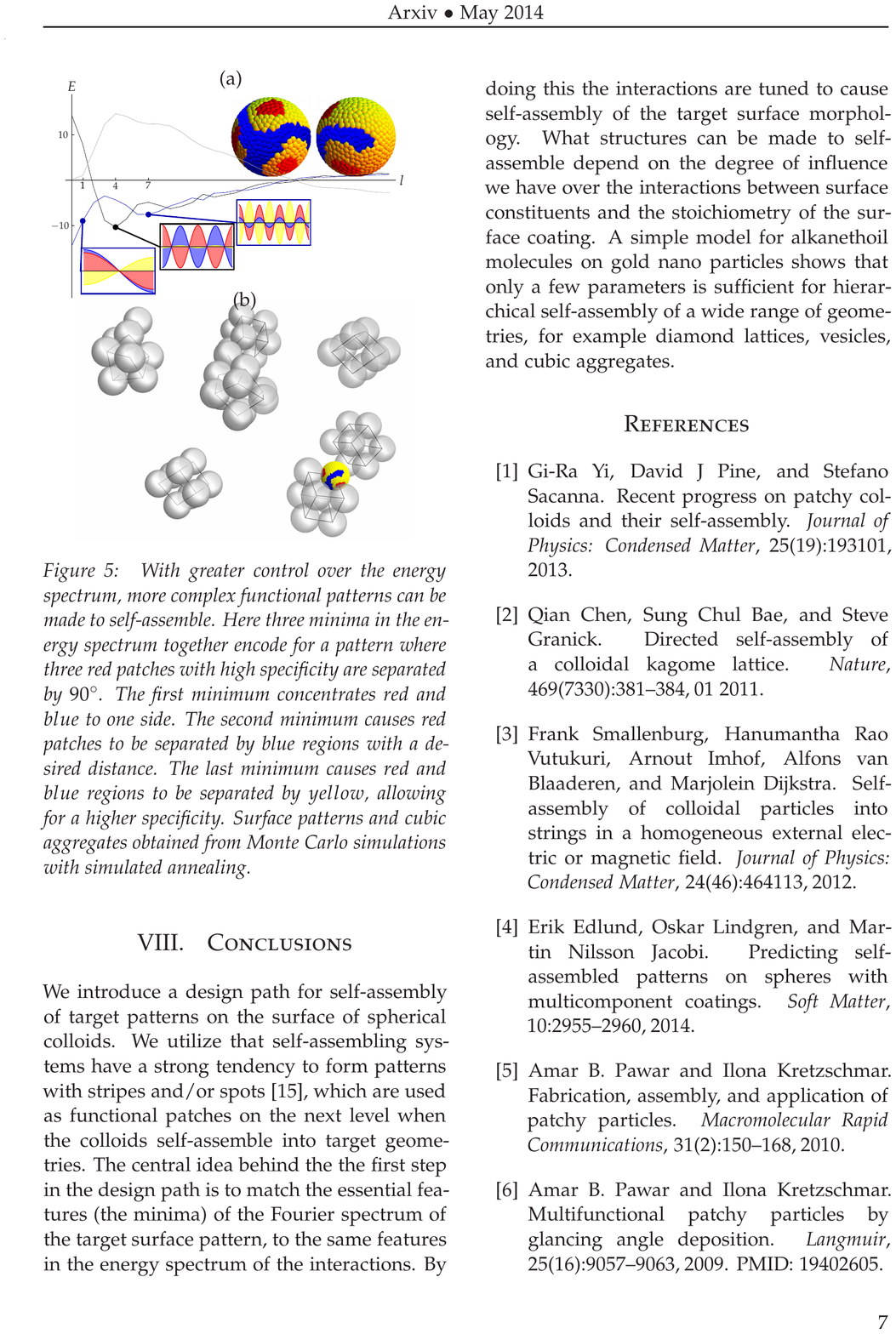}
\caption{ \label{cubeFig} With greater control over the energy spectrum, more complex functional patterns can be made to self-assemble. Here three minima in the energy spectrum together encode for a pattern where three red patches with high specificity are separated by $90^\circ$. The first minimum concentrates $red$ and $blue$ to one side. The second minimum causes red patches to be separated by blue regions with a desired distance. The last minimum causes $red$ and $blue$ regions to be separated by $yellow$, allowing for a higher specificity. Surface patterns and cubic aggregates obtained from Monte Carlo simulations with simulated annealing.}
\end{figure}

All three features are present for the alkanethiol-on-gold model if we choose $\epsilon = 0.08$, $L_{blue} = 1.1$, $L_{yellow} = 0$ and $L_{red} = 0.66$ radii. Choosing a stoichiometry of $21\%$ blue particles, $8\%$ red particles and $71\%$ yellow particles yields patches sufficiently small and with correct spacing, see the energy spectrum in the inset in Fig.~\ref{cubeFig}. The spacing and patch size can be adjusted to some degree by slight changes in the stoichiometry.
We verify the design by a Monte Carlo simulation of  a colloidal suspension. In Fig.~\ref{cubeFig} we see a low energy configuration where several cubes have self-assembled.% \obs{Discuss meta stability?}

%%%%%%%%%%%%%%%%%%%%%%%%%%%%%%%%%%%%%%%%%%%%%%%%%%%%%%%%%%%%%%%%
\section{Conclusions}

%In this paper we show examples of how limited control over interactions between surface constituents on coated nanoparticles is sufficient to cause hierarchical self-assembly of patterns, first on the surface of the nanoparticles then of the nanoparticles themselves. The design path utilises the fact that self-assembling systems dominated by long range isotropic interactions have a strong tendency to form stripes or spots like patterns CITE. The interactions ultimately only sets the length scale on which segregation between different constituents occurs at CITE. Fourier analysis of desired functional patterns provides both which length scales segregation needs to occur at as well as which length scales a given set of interactions will cause segregation at. What nanoparticle structures that can be made to self-assemble depends on to which degree we can influence the interactions between surface constituents and the stoichiometry of the surface coating. A simple model for alkanethiol molecules on gold nanoparticles show that only a few parameters is sufficient for hierarchical self-assembly of a wide range of geometries, from diamond patterns to cubic aggregates.

We introduce a design path for self-assembly of target patterns on the surface of spherical colloids. 
We utilize that self-assembling systems have a strong tendency to form patterns with stripes and/or spots \cite{Edlund_stripes_2010}, which are used as functional patches on the next level when the colloids self-assemble into target geometries.
The central idea behind the the first step in the design path is to match the essential features (the minima) of the energy spectrum of the interactions, to the maxima in the Fourier spectrum of the target surface pattern. By doing this the interactions are tuned to cause self-assembly of the target surface morphology. 
 What structures can be made to self-assemble depend on the degree of influence we have over the interactions between surface constituents and the stoichiometry of the surface coating.
A simple model for alkanethiol molecules on gold nanoparticles shows that only a few parameters is sufficient for hierarchical self-assembly of a wide range of geometries, for example diamond lattices, vesicles, and cubic aggregates.

\bibliographystyle{unsrt} 
\bibliography{multiparticleDesign2}

\end{multicols}

\end{document}